\documentstyle[psfig]{mn}

\newif\ifAMStwofonts

\def\v{\vec{v}}
\def\u{\vec{u}}
\def\x{\vec{x}}
\def\k{\vec{k}}
\def\brho{\bar{\rho}}
\def\drho{\delta\rho}
\def\mpc{h^{-1}{\rm Mpc}}
\ifoldfss
  \ifCUPmtlplainloaded \else
    \NewTextAlphabet{textbfit} {cmbxti10} {}
    \NewTextAlphabet{textbfss} {cmssbx10} {}
    \NewMathAlphabet{mathbfit} {cmbxti10} {} 
    \NewMathAlphabet{mathbfss} {cmssbx10} {} 
  \fi
  \ifAMStwofonts
    \ifCUPmtlplainloaded \else
      \NewSymbolFont{upmath} {eurm10}
      \NewSymbolFont{AMSa} {msam10}
      \NewMathSymbol{\upi}     {0}{upmath}{19}
      \NewMathSymbol{\umu}     {0}{upmath}{16}
      \NewMathSymbol{\upartial}{0}{upmath}{40}
      \NewMathSymbol{\leqslant}{3}{AMSa}{36}
      \NewMathSymbol{\geqslant}{3}{AMSa}{3E}

      \let\geq=\geqslant 
    \fi
  \fi
\fi 

\ifnfssone
  \newmathalphabet{\mathit}
  \addtoversion{normal}{\mathit}{cmr}{m}{it}
  \addtoversion{bold}{\mathit}{cmr}{bx}{it}
  \newmathalphabet{\mathbfit} 
  \addtoversion{normal}{\mathbfit}{cmr}{bx}{it}
  \addtoversion{bold}{\mathbfit}{cmr}{bx}{it}
  \newmathalphabet{\mathbfss} 
  \addtoversion{normal}{\mathbfss}{cmss}{bx}{n}
  \addtoversion{bold}{\mathbfss}{cmss}{bx}{n}
  \ifAMStwofonts
    \ifCUPmtlplainloaded \else
      \UseAMStwoboldmath
      \makeatletter
      \new@mathgroup\upmath@group
      \define@mathgroup\mv@normal\upmath@group{eur}{m}{n}
      \define@mathgroup\mv@bold\upmath@group{eur}{b}{n}
      \edef\UPM{\hexnumber\upmath@group}
      \new@mathgroup\amsa@group
      \define@mathgroup\mv@normal\amsa@group{msa}{m}{n}
      \define@mathgroup\mv@bold\amsa@group{msa}{m}{n}
      \edef\AMSa{\hexnumber\amsa@group}
      \makeatother
      \mathchardef\upi="0\UPM19
      \mathchardef\umu="0\UPM16
      \mathchardef\upartial="0\UPM40
      \mathchardef\leqslant="3\AMSa36
      \mathchardef\geqslant="3\AMSa3E

      \let\geq=\geqslant 
    \fi
  \fi
\fi 

\ifnfsstwo
  \DeclareMathAlphabet{\mathbfit}{OT1}{cmr}{bx}{it}
  \SetMathAlphabet\mathbfit{bold}{OT1}{cmr}{bx}{it}
  \DeclareMathAlphabet{\mathbfss}{OT1}{cmss}{bx}{n}
  \SetMathAlphabet\mathbfss{bold}{OT1}{cmss}{bx}{n}
  \ifAMStwofonts
    \ifCUPmtlplainloaded \else
      \DeclareSymbolFont{UPM}{U}{eur}{m}{n}
      \SetSymbolFont{UPM}{bold}{U}{eur}{b}{n}
      \DeclareSymbolFont{AMSa}{U}{msa}{m}{n}
      \DeclareMathSymbol{\upi}{0}{UPM}{"19}
      \DeclareMathSymbol{\umu}{0}{UPM}{"16}
      \DeclareMathSymbol{\upartial}{0}{UPM}{"40}
      \DeclareMathSymbol{\leqslant}{3}{AMSa}{"36}
      \DeclareMathSymbol{\geqslant}{3}{AMSa}{"3E}

      \let\geq=\geqslant 
    \fi
  \fi
\fi 

\ifCUPmtlplainloaded \else
  \ifAMStwofonts \else 
    \def\upi{\pi}
    \def\umu{\mu}
    \def\upartial{\partial}
  \fi
\fi
\title{ Spherical Infall Model  in a Cosmological Background
  Density Field} 
\author[A. Taruya, J. Soda]
        {A. Taruya$^1$, J. Soda$^2$ \\
        $^1$Research Center for the Early Universe (RESCEU)
            School of Science, University of Tokyo, 
            Tokyo, 113-0033, Japan\\
        $^2$Department of Fundamental Sciences, FIHS, Kyoto University, 
            Kyoto, 606-8501, Japan}
\date{\today}
\pagerange{\pageref{firstpage}--\pageref{lastpage}}
\pubyear{2000}

\begin{document}

\maketitle

\label{firstpage}

%
%
%
%
%
%
%
%
%
%
%
\begin{abstract}
We discuss the influence of the cosmological background density field  
 on the spherical infall model. 
The spherical infall model has been used in the Press-Schechter 
formalism to evaluate the number abundance of clusters of galaxies,  
as well as to determine the density parameter of the universe 
from the infalling flow. Therefore, the understanding of 
collapse dynamics play a key role for extracting the 
cosmological information. Here, we consider 
the modified version of the spherical infall model. 
We derive the mean field equations from the Newtonian fluid 
equations, in which the influence of cosmological background inhomogeneity 
 is incorporated into the averaged quantities as the 
{\it backreaction}. By calculating the averaged quantities
 explicitly, we obtain the simple expressions and find that 
in case of the scale-free power spectrum, the density  
fluctuations with the negative spectral index make the infalling 
velocities slow. This suggests that we underestimate the 
density parameter $\Omega$ when using the simple spherical infall model. 
In cases with the index $n>0$, the effect of background 
inhomogeneity could be negligible and the spherical infall model 
becomes the good approximation for the infalling flows. 
We also present a realistic example with the cold dark 
matter power spectrum. There, the anisotropic random velocity 
leads to slowing down the mean infalling velocities. 
\end{abstract}
%
%
%
%
%
%
%
%
%
%
%
%
%
%
%
%
%
\section{Introduction}
\label{sec: intro}
%
%
%
%
In the standard scenario of the cosmic structure formation, 
      the density peaks in the large-scale inhomogeneities 
      grow due to the gravitational instability and 
      they  experience the gravitational collapse,   
      which finally lead to the virialized objects such 
      as the clusters of galaxies or the dark matter halos. 

Modeling such  non-linear objects plays 
      an important role for probing the large scale structure. 
      After the seminal paper by Gunn \& Gott (1972), the spherical 
      infall model has been extensively used in the literature.  
The non-linear system mimicking the density peak 
      is represented by the spherical overdensity 
      with the radius $R$, which 
      obeys the simple dynamics (Peebles 1980)
\begin{equation}
 \ddot{R}+\frac{GM}{R^2}=0,
  \label{EOM-SCM}
\end{equation}
      where $M$ is the mass of the bound object. The density 
      contrast of this non-linear object is defined by 
      $\delta=(a/R)^3-1$, where $a$ is the expansion factor of the 
      universe. From (\ref{EOM-SCM}), 
      we can estimate the collapse time $t_c$, which gives the 
      dynamical time scale of the virialization. In terms of the 
      linearly extrapolated density contrast,  
      we have the value $\delta_c\equiv\delta(t_c)\simeq1.68$ 
      in the Einstein-de Sitter universe (for the low density universe
      or the flat universe with the non-vanishing cosmological constant, 
      see e.g, Eke, Coles \& Frenk 1996). This is frequently used in the 
      Press-Schechter mass function to evaluate the number abundance 
      of cluster of galaxies (Press \& Schechter 1974, 
      Eke, Coles \& Frenk 1996, Kitayama \& Suto 1997). 

      Further, the spherical infall model can also be applied to 
      estimate the density parameter of our universe, although 
      the other efficient method such as POTENT has recently been 
      available (e.g, Bertschinger \& Dekel 1989, see also 
      the review by Strauss \& Willick 1995). 
      The observation of the infalling velocity $V_{\rm inf}$  
      can be compared to the expansion rate $\dot{R}/R$ subtracting 
      the uniform Hubble flows. 
      Characterizing the infalling velocity $V_{\rm inf}$ as 
      the function of mean overdensity $\bar{\delta}$ 
      within the sphere centered 
      on the core of cluster, we can determine 
      $\Omega$ from the $\bar{\delta}$-$V_{\rm inf}$ relation 
      (Davis \& Huchra 1982, Yahil 1985, Dressler 1988). 

As we know, the spherical infall model is the ideal system. 
      In real universe, the density peak cannot evolve locally. 
      The cosmological background density field affects the 
      dynamical evolution and  
      the collapsed time of the density peak. This might lead to the 
      incorrect prediction for the low mass objects using 
      the Press-Schechter formalism (Jing 1998, Sheth, Mo \& Tormen 1999). 
      As for the infall velocity, the tidal effect distorts the 
      flow field around the core of cluster, which becomes 
      the source of the underestimation of 
      the density parameter 
      (Villumsen \& Davis 1986).  

Therefore, the understanding of the influence of cosmological background 
      density field on the bound objects is crucial to clarify 
      the cosmic structure formation. Several authors treat 
      this issue and consider the modification 
      of the spherical infall model taking into account the background 
      density inhomogeneity (Hoffman 1986, Bernardeau 1994, 
      Eisenstein \& Loeb 1995). The validity of the spherical 
      infall model and the 
      estimation of the collapsed time scale have been mainly discussed.  
      Another author evaluates the weakly non-linear correction of 
      the background density field using the conditional probability 
      distribution function around the overdensity 
      and studies the influence of the non-linear density 
      inhomogeneities to the dark matter density profiles (\L okas 1998).  
      

In this paper, we investigate the influences of the cosmic background 
      density field on the spherical infall model using the different
      approach from the previous works. 
      Focusing on the mean expansion around the bound system, 
      the Gauss' law implies that  the 
      interior of the bound object can be regarded as the 
      homogeneous overdensity by taking the averaging 
      procedure. Hence, we will treat the averaged dynamics of 
      the inhomogeneous gravitational collapse. 
      The influence of cosmic density fields on 
      the infalling velocity is incorporated into 
      the averaged quantities as 
      the {\it backreaction} of the growing inhomogeneities. 
      Importantly, the modified dynamics 
      can be non-local. In addition to the mass $M$ of the bound 
      object, the evolution depends on the statistical property of
      the density fluctuations. 

In section \ref{sec: dynamics}, we first argue how the background 
density field modifies the non-linear dynamics (\ref{EOM-SCM}).  
Then, we derive the evolution equations for overdensity by 
averaging the inhomogeneous Newtonian gravity. The modified dynamics 
can contain the averaged quantities characterizing the backreaction of 
the cosmic density inhomogeneity. We explicitly calculate these 
quantities in the Appendix A and obtain the simple expressions. 
The main results in this paper are equations (\ref{average-1}) and 
(\ref{average-2}). Using these expressions, we further discuss the 
influence of backreaction effect on the infalling velocity  
in section \ref{sec: estimation}. Final section \ref{sec: conclusion} 
is devoted to the summary and discussions.  
%
%
%
%
%
%
\section[]{Averaged dynamics of spherical infall model}
\label{sec: dynamics}
%
%
%
%
%
%
\subsection[]{The backreaction of cosmic density field}
\label{subsec: derivation}
%
%
%
%
%
%
%
%
We shall start with  the basic equations 
for Newtonian dust fluid:  
  \begin{eqnarray}
   && \frac{\partial \rho}{\partial \tau}+\vec{\nabla}_X(\rho\v)=0, 
    \nonumber
\\
   && \frac{\partial \v}{\partial
   \tau}+(\v\cdot\vec{\nabla}_X)\v=-\vec{\nabla}_X\phi
    \label{eq: start-up}
\\
   && \nabla_X^2\phi=4\pi G \rho,
   \nonumber
  \end{eqnarray}
where $\tau$ is the proper time of the dust fluid and 
$\vec{\nabla}_X$ is the derivative with respect to the proper 
distance $\vec{X}$ from some chosen origin.  

The non-linear dynamics of spherical overdensity 
       (\ref{EOM-SCM}) is embedded in the above system. 
       Consider the homogeneous infalling flow, in which the 
       quantities $\rho$, $\vec{v}$ and $\phi$ are 
       given by 
  \begin{eqnarray}
    \rho&=&\bar{\rho}(\tau)
    \nonumber
\\
   \v&=& {\cal H}(\tau)\vec{X} 
    ~;~~~~~{\cal H}=\frac{\dot{R}}{R},  
    \label{ansatz}
    \nonumber
\\
   \phi&=&\frac{2}{3}\pi G~\brho(\tau)~|\vec{X}|^2
    \nonumber
  \end{eqnarray}
       where $R$ is the proper radius of the spherical overdensity. 
We then introduce the comoving frame $(t, \vec{x})$ defined as 
\begin{equation}
      \tau=t,~~~~~~~\vec{X}=R(t)\x.  
\end{equation} 
Substitution of (\ref{ansatz}) into (\ref{eq: start-up}) yields 
\begin{eqnarray}
\dot{\brho}+3{\cal H}\brho=0, ~~~~~
3\dot{\cal H} + 3{\cal H}+4\pi G \brho=0, 
\label{SIM}
\end{eqnarray}
where $(\dot{~})$ denotes the derivative with respect to the time $t$. 
Because the first equation implies the mass conservation 
$M=(4\pi/3)\brho R^3$, the homogeneous system 
(\ref{SIM}) is equivalent to the dynamics (\ref{EOM-SCM}). 

When we take into account the cosmic background density field, 
       the dynamics of overdensity affects 
       the evolution of background inhomogeneity and 
       the growth of inhomogeneities induces the tidal force, 
       which distorts the homogeneous evolution of the infalling flows. 
       The situation we now investigate is 
       that the background fluctuation 
       is not so large and the dynamics of overdensity is dominated by 
       the radial motion, that is, the dynamics could be 
       almost the spherically symmetric collapse. In this case, we can 
       approximately treat the radial infalling flows 
       as the homogeneous system and the evolution   
       of the overdensity is solely affected by the small perturbation 
       of the background density fields.
 
Therefore, the problem is reduced to a classic 
       issue of the {\it backreaction}. That is, 
       together with the quantities $(\brho, {\cal H})$, 
       the dynamics of the homogeneous overdensity can be 
       determined by the backreaction of background inhomogeneities 
       through the non-linear interaction.  
       Phenomenologically, the backreaction effect can be mimicked 
       by adding the non-vanishing terms characterizing the 
       density inhomogeneity 
       in the right hand side of equations (\ref{SIM}).   
Here, we consider the self-consistent treatment based on 
the equations (\ref{eq: start-up}). We will derive the 
dynamics of infalling flows including the backreaction effect. 
       Since the evolution of overdensity is affected by 
       the spatial randomness of the cosmic inhomogeneity,  
       it should be better to investigate the mean dynamics of the 
       overdensity, which can be derived 
       from the system (\ref{eq: start-up}) taking the spatial 
       averaging. 

Let us divide the quantities 
       $\rho$, $\vec{v}$ and $\phi$ into the 
       homogeneous part and the fluctuating part with the zero mean:
  \begin{eqnarray} 
    \label{seperate}
    \rho&=&\bar{\rho}(\tau)+ \drho(\tau, \vec{X}),
    \nonumber\\
   \v&=& {\cal H}(\tau)\vec{X} + \u(\tau, \vec{X}),  
    \label{division} 
    \\
   \phi&=& \frac{2}{3}\pi G~\brho(\tau)~|\vec{X}|^2 + \Phi(\tau, \vec{X}).
    \nonumber
  \end{eqnarray}
%
 Substituting (\ref{division}) into (\ref{eq: start-up}),  
   the equation of continuity becomes 
\begin{eqnarray}
&& \left(\dot{\brho}+3{\cal H}\brho\right)+
  \left(\frac{\partial \delta\rho}{\partial t}+3{\cal H}\delta\rho+
  \frac{\brho}{R}\vec{\nabla}_x\vec{u}\right)
  \nonumber \\
&& ~~~~~~~~~~~~~~~~~~~~~~~~~~~~~~~~~~~~~~~~=-\frac{1}{R}
  \vec{\nabla}_x(\delta\rho~\vec{u}).
  \label{reduced1}
\end{eqnarray}
For the Euler equation, taking the spatial divergence yields 
\begin{eqnarray}
&& \left(3\dot{\cal {H}}+3{\cal H}^2+ 4\pi G \brho\right)+
 \frac{1}{R}\vec{\nabla}_x\left(\frac{\partial \vec{u}}{\partial t}+
        {\cal H}\vec{u}+\frac{1}{R}\vec{\nabla}_x\Phi \right)
  \nonumber\\
&& ~~~~~~~~~~~~~~~~~~~~~~~~~~~~~~
 =~-\frac{1}{R^2}\nabla_x\left[(\vec{u}\cdot\vec{\nabla}_x)\vec{u}\right].
  \label{reduced2}
\end{eqnarray}
In equations (\ref{reduced1}) and (\ref{reduced2}),  
the non-linear interaction of the background inhomogeneities are 
expressed in the right hand side of these equations.

The system (\ref{reduced1}) and (\ref{reduced2}) can be 
       divided into the homogeneous part and the fluctuating part 
       by taking the spatial average over the comoving volume 
       $V_r=(4\pi/3)r_0^3$, where the radius $r_0$ denotes the comoving 
       size of the characteristic overdensity. The spatial averaging 
       is defined as 
\begin{equation}
 \langle\hspace*{-0.7mm}\langle\cdots\rangle\hspace*{-0.7mm}\rangle
=\frac{1}{V_r}\int_{V_r}d^3x~\langle\cdots\rangle, 
\label{average}
\end{equation}
       where $\langle\cdots\rangle$ denotes the ensemble average taken 
       over the random fluctuations $\delta\rho$ and $\vec{u}$. 
%
Then, we obtain
\begin{eqnarray}
&&\dot{\brho}+3{\cal H}\brho=
-\frac{1}{R}\langle\hspace*{-0.7mm}\langle\vec{\nabla}_x\left[\drho~\u\right]
\rangle\hspace*{-0.7mm}\rangle , 
\label{rho-evolution}
\\
&&  \frac{\ddot{R}}{R}+\frac{4\pi G}{3}\brho=
  -\frac{1}{3R^2}\langle\hspace*{-0.7mm}\langle\vec{\nabla}_x\left
[(\u\cdot\vec{\nabla}_x)\u\right]
\rangle\hspace*{-0.7mm}\rangle
\label{R-evolution} 
\end{eqnarray}
for the homogeneous part and  
%
\begin{eqnarray}
&&  \frac{\partial\drho}{\partial t}+3{\cal H}\drho+
\frac{1}{R}\vec{\nabla}_x\u=0,
\nonumber \\
&&  \frac{\partial \u}{\partial t}+{\cal H}\u
=\frac{1}{R}\vec{\nabla}_x\Phi=0,
\label{perturbation-evolution} \\
&& \nabla_x^2\Phi=4\pi G R^2 \drho, 
\nonumber
\end{eqnarray}
for the fluctuating part with the zero mean. 

The results (\ref{rho-evolution}), (\ref{R-evolution}) and 
       (\ref{perturbation-evolution}) are the desirable forms 
       for our purpose. The mean dynamics of overdensity is affected 
       by the the averaged quantities 
       $\langle\hspace*{-0.7mm}\langle\cdots\rangle\hspace*{-0.7mm}\rangle$   
       corresponding to the backreaction of the cosmic density 
       inhomogeneities. These quantities are evaluated 
       self-consistently by solving the evolution equations 
       (\ref{perturbation-evolution}) 
       under the knowledge of homogeneous quantities $\brho$ and $R$. 
%
%
%
%
%
%
\subsection[]{The averaged quantities}
\label{subsec: averaging}
%
%
%
%
%
%
We proceed to analyze the mean dynamics 
       (\ref{rho-evolution}) and (\ref{R-evolution}).  
       In general relativity, 
       the quantitative analysis of the backreaction terms 
       is still hampered by a more delicate issue of 
       the gauge conditions even after deriving the basic 
       equations (Buchert 1999). However, in the Newton gravity, 
       we are not worried about the evaluation of the averaged 
       quantities. 
       Let us separate the variables $\drho$, $\u$ and $\Phi$ into the 
       the time-dependent part and the spatial random part: 
\begin{eqnarray}
 \delta\rho &=& ~D(t) ~ \hat{\delta}(x), 
\nonumber \\
 \vec{u} &=& ~G~R(t)V(t) ~ \vec{\nabla}_x\hat{\Delta}(x), 
\label{separated-var} \\
\Phi &=&  ~G~R^2(t)D(t) ~ \hat{\Delta}(x), 
\nonumber 
\end{eqnarray}
where the variables with $(\hat{~})$ denote the random field. 
The ensemble average is taken with respect to these variables.  
Then, the evolution equations (\ref{rho-evolution}), (\ref{R-evolution}) 
and (\ref{perturbation-evolution}) become the 
four set of ordinary differential equation: 
\begin{eqnarray}
 \dot{\brho}~~+~~3{\cal H}\brho&=&-G(DV)~{\bf C}_1, 
\label{evolution1} \\
\frac{\ddot{R}}{R}~+\frac{4\pi G}{3}\brho&=&-\frac{G^2}{3}V^2~{\bf C}_2, 
\label{evolution2} \\
 \dot{D}~+~3{\cal H}D&=&4\pi G~\brho~V, 
\label{evolution3} \\
 \dot{V}~+~2{\cal H}V&=&D, 
\label{evolution4}
\end{eqnarray}
where 
\begin{eqnarray}
 {\bf C}_1&=&\frac{\langle\hspace*{-0.7mm}
  \langle\vec{\nabla}_x\left[\drho~\u~\right]\rangle\hspace*{-0.7mm}\rangle}
  {G~(RVD)},  
\nonumber 
\\
 {\bf C}_2&=&\frac{\langle\hspace*{-0.7mm}\langle\vec{\nabla}_x\left
[(\vec{u}\cdot\vec{\nabla}_x)\u\right]
\rangle\hspace*{-0.7mm}\rangle}{ G^2(RV)^2}
\nonumber 
\end{eqnarray}
are merely the numerical constants. 
Provided the statistical feature of the spatial inhomogeneities
$\hat{\delta}$ and $\hat{\Delta}$, we can evaluate ${\bf C}_1$ and 
${\bf C}_2$. Notice that both $\hat{\delta}$ and $\hat{\Delta}$ 
are related through the Poisson equation:
\begin{equation}
 \nabla^2_x~\hat{\Delta}=~-4\pi~\hat{\delta}.
  \label{Poisson} 
\end{equation}
Thus, the non-local tidal effect induced by the cosmic density fields 
can be treated in our prescription. 

To evaluate the averaged quantities ${\bf C}_1,~{\bf C}_2$, 
       it is sufficient to know the second order statistics 
       of the density inhomogeneities, i.e,  
       the power spectrum $P(k)$. Usually, $P(k)$ is given by 
\begin{equation}
 \langle\hat{\delta}(\k)\hat{\delta}(\k')\rangle
  =(2\pi)^3\delta_D(\k+\k')~P(k) 
  \label{P(k)}
\end{equation}
       in the Fourier representation. 
However, the naive computation using the above definition 
       gives the monopole term, i.e, $l=0$ mode of the 
       spherical harmonic expansion for the fluctuations. 
       In our approach, the coherent radial motion 
       arising from the monopole contribution has already been 
       {\it renormalized} in the homogeneous system, that is,   
       the fluctuations defined in (\ref{division}) 
       have only the higher multipole $l\geq1$. 
       This means that the alternative 
       definition subtracting the monopole term 
\begin{equation}
  \langle\hat{\delta}(\k)\hat{\delta}(\k')\rangle
  =(2\pi)^3~P(k) \left[\delta_D(\k+\k')-\frac{1}{4\pi k^2}\delta_D(k-k')\right]
\label{modified-P(k)}
\end{equation}
       should be used in our formalism. 

The calculation using (\ref{modified-P(k)}) is essential to 
       obtain the non-vanishing averaged quantities. 
       In the Appendix A, the details of calculation for averaged quantities 
       are described. Here, we present the final expressions: 
\begin{eqnarray}
{\bf C}_1&=&
\frac{8}{V_r}\left\{\xi_0(r_0)-\xi_1(r_0)\right\}. 
\label{average-1}
\\
{\bf C}_2&=&\frac{32\pi}{3~V_r}
\left\{-\xi_0(r_0)+3\xi_1(r_0)-2\xi_2(r_0)\right\}, 
\label{average-2}
\end{eqnarray}
The function $\xi_l(r_0)$ denotes the statistical quantity 
 given by 
\begin{equation}
 \xi_l(r_0)\equiv~\int_0^{r_0}dr~\int_0^{\infty} dk~(kr)^2 ~
  P(k)\left[j_l(kr)\right]^2, 
\label{xi}
\end{equation}
where $j_l(x)$ is the spherical Bessel function for $l$-th order. 

Equations (\ref{average-1}) and (\ref{average-2}) are the main results 
of this paper. From these expressions, we immediately 
see that the backreaction effect becomes negligible on large scales 
due to the 
suppression factor $1/V_r$. This is correct in the cases with the 
power-law spectrum $P(k)\propto k^n$ within the range $-3<n<1$. 
In the limit $r_0\to\infty$, 
the dynamics (\ref{rho-evolution}) and (\ref{R-evolution}) 
can be well-approximated by the spherical 
infall model (\ref{EOM-SCM}). 
%
%
%
%
%
%
%
%
%
%
%
%
%
%
%
%
%
%
%
%
\begin{figure}
\begin{center}
\leavevmode
\psfig{file=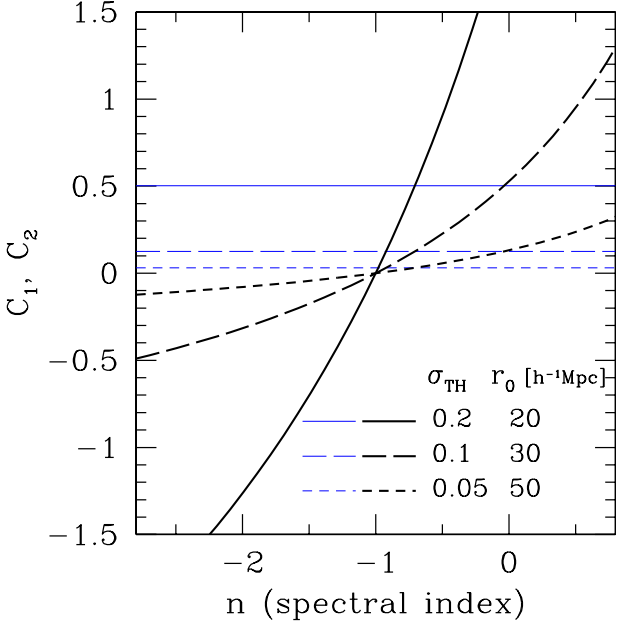,width=9.5cm}
\end{center}
\caption{The averaging quantities ${\bf C}_{1,2}$ as the function of 
$n$ for a power-law spectrum $P(k)~\propto~k^n.$ The thin and thick 
 lines respectively denote the quantities ${\bf C}_1$ and ${\bf C}_2$,  
which are evaluated from (\protect\ref{average-1}) and (\protect\ref{average-2}) 
 using the expressions (\protect\ref{xi_l}) and (\protect\ref{normalize-A}).
}
\end{figure}
%
%
%
%
%
%
%
%
%
%
%
%
%
\begin{figure}
\begin{center}
\leavevmode
\psfig{file=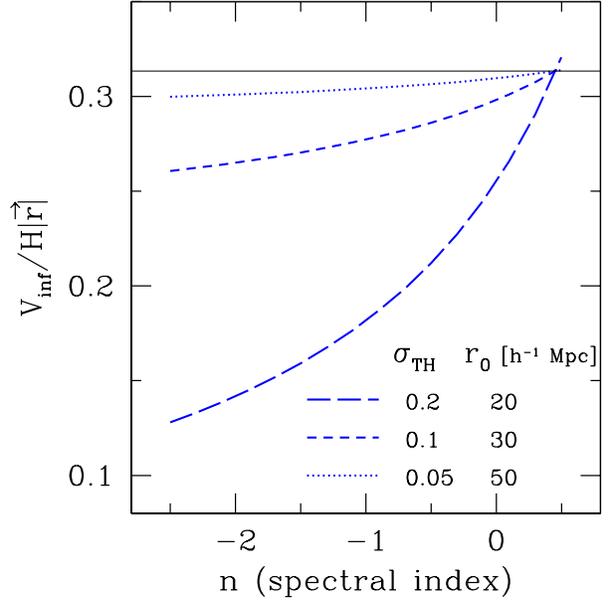,width=9.3cm}
\end{center}
\caption{The ratio of the infalling velocity to the Hubble flow as the 
function of the spectral index $n$ ({\it long-dashed}, 
{\it short-dashed}, {\it dotted}, {\it dot-dashed-lines}). 
By solving the mean dynamics 
(\protect\ref{evolution1})-(\protect\ref{evolution4}), 
the quantity given by
(\protect\ref{V_infall}) is evaluated at the time $a_f/a_i=4$ in the 
Einstein-de Sitter universe. We choose the initial density 
contrast $\bar{\delta}=0.3$. The initial condition for the 
fluctuating part are described by the linear growing mode of 
perturbations neglecting the backreaction effect. 
The results are compared to the homogeneous infalling flow obtained from 
 the simple dynamics (\protect\ref{EOM-SCM})({\it thin-solid line}).
}
\end{figure}
%
%
%
%
%
%
%
%
%
%
%
%
\section[]{Effects on the infalling velocity }
\label{sec: estimation}
%
%
%
%
Provided the power spectrum $P(k)$ of the background 
  density fluctuations, we can investigate the mean infalling flows 
  around the density peak. For the present purpose, it is useful to 
  consider the scale-free spectrum in more detail: 
  \begin{equation}
    P(k)=A~k^n. 
  \end{equation}
Then the quantity $\xi_l(r_0)$ can be expressed in terms of the 
gamma functions. We have 
\begin{equation}
 \xi_l(r_0)=-A~r_0^{-n}~\frac{2^n\pi}{n}~
  \frac{\Gamma{(-n-1)}\Gamma{(l+(3+n)/2)}}
  {[\Gamma{(-n/2)}]^2\Gamma{(l+(1-n)/2)}}.
\label{xi_l}
\end{equation}
The power spectrum is related to the variance of density fluctuations 
$\sigma_{\scriptscriptstyle\rm TH}$, where the subscript
$_{\scriptscriptstyle\rm TH}$ means the variable 
with the top-hat smoothing. 
Identifying the 
smoothing radius with the comoving radius of overdensity $r_0$, 
the influence of the background inhomogeneities to the overdensity can be
quantified by $\sigma_{\scriptscriptstyle\rm TH}$. 
Then, the normalization factor $A$ is given by 
\begin{equation}
A=\frac{4\pi}{3}~r_0^{3+n}\sigma_{\scriptscriptstyle\rm TH}^2~
\frac{[\Gamma((2-n)/2)]^2\Gamma((3-n)/2)}{2^{n-1}\Gamma(1-n)\Gamma((3+n)/2)}.
\label{normalize-A}
\end{equation}
Thus, the averaged quantities can be 
characterized by the three parameters,  $r_0$, $n$ and 
$\sigma_{\scriptscriptstyle\rm TH}$. 

In Fig.1, we plot the averaged 
quantities as the function of the spectral index $n$. 
The typical parameters are respectively chosen as follows: 
$\sigma_{\scriptscriptstyle\rm TH}=0.2$ and $r_0=20~h^{-1}$ Mpc 
({\it solid line});    
$\sigma_{\scriptscriptstyle\rm TH}=0.1$ and $r_0=30~h^{-1}$ Mpc 
({\it long-dashed line}) ; 
$\sigma_{\scriptscriptstyle\rm TH}=0.05$ and 
$r_0=50~h^{-1}$ Mpc ({\it short-dashed line}). 
Fig.1 says that as the variance $\sigma_{\scriptscriptstyle\rm TH}$ 
decreases and/or the radius $r_0$ increases,  
the quantities ${\bf C}_1$ and ${\bf C}_2$ become negligible. 
The thick lines shows the variable ${\bf C}_2$. 
      Interestingly, ${\bf C}_2$ turns its signature. 
      The negative value is obtained at $n<-1$, while 
      we have the positive value ${\bf C}_2$ at $n>-1$. 
      By contrast, the quantity ${\bf C}_1$ denoted as 
      the thin lines becomes positive and almost constant over the range 
      $-2.8<n<0.8$.  

Fig.1 indicates that backreaction of the cosmic density inhomogeneities 
can prevent the spherical shell from infalling into the center of 
overdensity for the negative spectral index 
(see eqs.[\ref{evolution1}][\ref{evolution2}]). 
The effect becomes noticeable when $n<-1$.  
      The result can be interpreted as follows. The quantities ${\bf C}_2$ 
      is expressed in terms of the geometrical optics. 
      From (\ref{separated-var}), we can write the peculiar velocity 
      as $\partial_iu_j=(\theta/3)\delta_{ij}+\sigma_{ij}$, where 
      $\theta$ and $\sigma_{ij}$ are the expansion scalar and the 
      shear tensor, respectively. Then we have 
\begin{equation}
\langle\hspace*{-0.7mm}\langle\vec{\nabla}_x\left
[(\u\cdot\vec{\nabla}_x)\u\right]
\rangle\hspace*{-0.7mm}\rangle =
\langle\hspace*{-0.7mm}\langle
\frac{1}{3}\theta^2
\rangle\hspace*{-0.7mm}\rangle +
\langle\hspace*{-0.7mm}\langle
\sigma_{ij}\sigma^{ij}
\rangle\hspace*{-0.7mm}\rangle +
\langle\hspace*{-0.7mm}\langle
\vec{u}\cdot\vec{\nabla}\theta
\rangle\hspace*{-0.7mm}\rangle. 
\label{udevu}
\end{equation}
      The last term in the right-hand side of (\ref{udevu}) 
      means the flows of inertial frame.  
      It is easy to show that the first and the 
      second terms in the right-hand side of (\ref{udevu}) always 
      become positive. On the other hand, the last term can be written by 
\begin{equation}
\frac{ \langle\hspace*{-0.7mm}\langle
\vec{u}\cdot\vec{\nabla}\theta
\rangle\hspace*{-0.7mm}\rangle}{G^2(RV)^2}
=-\frac{32\pi}{V_r}\left[\sum_{l=0}(2l+1)\xi_l(r_0)-\xi_1(r_0)\right], 
\end{equation}      
      which is always negative. This means that the 
      anisotropic flows of cosmic density field can induce 
      the effective pressure 
      and this could dominate the expansion and the shear terms.  
      The effect becomes significant when the long wavelength 
      fluctuations have the large power, corresponding to the 
      small spectral index $n<-1$. 
      Accordingly, the background inhomogeneity apparently 
      prevents the collapse of the averaged overdensity.  

To see the backreaction effect more explicitly, 
      we analyze the infalling velocity. The flow field 
      around the spherical overdensity can be quantified by 
\begin{equation}
 \frac{V_{\rm inf}}{H|\vec{r}|}\equiv\frac{H-{\cal H}}{H}.
\label{V_infall}
\end{equation}
      The above quantity can be evaluated by solving the evolution equations 
      (\ref{evolution1})-(\ref{evolution4}).  

In Fig.2, we plot the ratio of the infalling velocity to the Hubble 
      flow as the function of the spectral index. Here, 
      the cosmic expansion is assumed to be that of the  
      Einstein-de Sitter universe, $a=(t/t_i)^{2/3}$. 
      The initial velocity flow ${\cal H}$ of the overdensity is obtained 
      by equating it with the Hubble expansion. As for the initial 
      overdensity, $\bar{\rho}$ can be characterized by the 
      density contrast $\bar{\delta}=\bar{\rho}/\rho_b-1$, 
      where $\rho_b$ is the homogeneous density of the universe. 
      We choose the initial density contrast 
      as $\bar{\delta}=0.3$. On the other hand, the evolution of 
      the fluctuating part $D$ and $V$ can be initially  
      described by the linear growing mode neglecting the backreaction effect. 
      Then, we solve the evolution equations by varying the 
      spectral index $n$. 
      The results are depicted when the final expansion 
      factor $a_f$ becomes four times larger than the initial 
      expansion factor $a_i$, i.e, $a_f/a_i=4$, within the 
      validity of the approximation neglecting the higher order 
      perturbations. 

In Fig.2, we also plot the ratio $V_{\rm inf}/H|\vec{r}|$ 
      evaluated from the 
      spherical infall model (\ref{EOM-SCM}) ({\it thin-solid line}).  
      Clearly, the deviation from the spherical infall 
      model becomes significant for the small radius of the 
      overdensity and the large variance of the fluctuations 
      ({\it thick-long-dashed, thick-short-dashed lines}). 
      The cosmic background 
      inhomogeneity can weaken the infalling velocity. 
      This effect leads to the overestimation of the cosmic 
      expansion when we compare the observation of the 
      infalling velocity with the spherical infall model.  
      Consequently, we would prefer the low density universe. 
      However, the deviation becomes small as the spectral index 
      increases. The backreaction effect 
      becomes negligible and the infalling velocity can be 
      well-described by the spherical infall model. 
      At the time $a_f/a_i=4$, the index $n\approx0.5$ is 
      marginal, which is rather larger value than $n=-1$. 

As another example, a realistic case with the 
cold dark matter (CDM) spectrum is examined. In Fig.3, we specifically 
consider the standard CDM, i.e, $\Omega_0=1$ and $\lambda_0=0$. 
The fitting form of the power spectrum given by Bardeen et al. (1986) 
is used and normalized by the top-hat fluctuation amplitude at 
$8h^{-1}$Mpc, $\sigma_8=0.6$, according to the cluster abundance 
(Kitayama \& Suto 1997). 
The Hubble parameter is given by $H_0=100\,h\,{\rm km}\, s^{-1}/{\rm Mpc}$ 
with $h=0.5$.  In Fig.3, the present values of $C_{1,2}$ 
(upper-panel) and $V_{\rm inf}/H|\vec{r}|$ (lower-panel) are depicted 
as the function of the averaging scale $r_0$. The initial conditions 
for the lower-panel are set at $z=5$, similar to the Fig.2. 
The figure states that both quantities $C_1$ and $C_2$ become positive  
on large scales $r_0 \geq 30\mpc$ and the resulting ratio 
$V_{\rm inf}/H|\vec{r}|$ takes the slightly lower value, compared to the 
original spherical infall model ({\it thin-solid line} in lower-panel). 
This is because the dominant contribution of the CDM spectrum to the 
averaging quantities is the low-$k$ part with the effective spectral index 
$-1<n_{\rm eff}<0$, where $n_{\rm eff}\equiv d\log{P(k)}/d\log{k}$ 
(see Fig.1). The behavior of infalling velocity is thus understood from Fig.2. 
On the other hand, the deviation of $V_{\rm inf}/H|\vec{r}|$ 
from the spherical infall model becomes large as decreasing the 
averaging size $r_0$. The quantity $C_2$ turns its signature 
at the relatively small size, $r_0\simeq9\mpc$. 
In any cases, the present model indicates that the method using 
infalling velocity still leads to the underestimation of the 
density parameter in real universe. 
%
%
%
%
%
%
%
%
%
%
%
%
\begin{figure}
\begin{center}
\leavevmode
\psfig{file=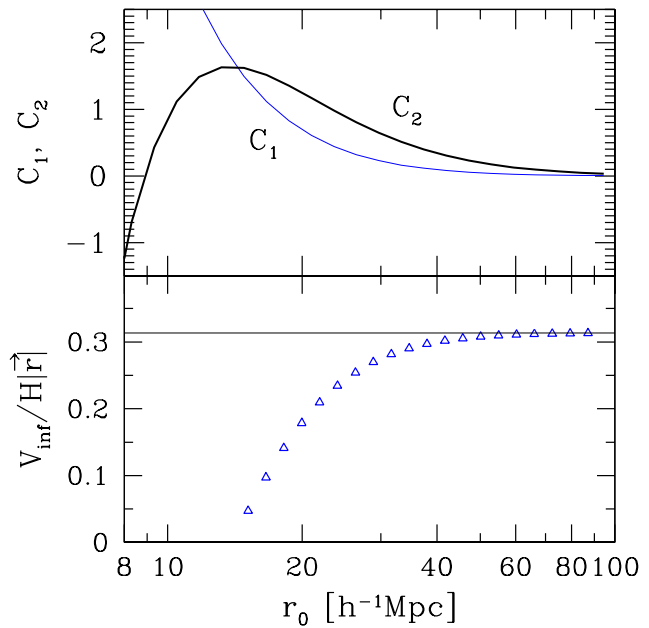,width=9.3cm}
\end{center}
\caption{Scale-dependence of the averaging quantities 
${\bf C}_{1,2}$ (upper-panel) and the ratio of the infalling velocity 
to the Hubble flow (lower panel) in the case of the standard CDM  
model. In both panels, we used the power spectrum $P(k)$ given by 
Bardeen et al. (1986) normalizing at the present epoch 
with the parameters $(\Omega_0, \lambda_0, h,
\sigma_8)=(1.0,0.0,0.5,0.6)$.    
Upper-panel: the averaging quantities $C_1$ ({\it thin-solid line}) and 
$C_2$ ({\it thick-solid line}) evaluated at present ; 
Lower-panel: the present values of the ratio $V_{\rm inf}/H|\vec{r}|$ 
evolved from $z=5$ for the original spherical infall model 
({\it thin-solid line}) and our model ({\it open triangles}).  
}
\end{figure}
%
%
%
%
%
%
%
%
%
%
%
\section{Discussion and conclusion}
\label{sec: conclusion}
%
%
%
%
%
%
%
       We have analyzed the mean dynamics of the 
       spherical infall by taking into account the effects of 
       cosmic background density field self-consistently. 
       After deriving the evolution equations, we 
       have rigorously computed the averaged quantities appearing in 
       the equation of the homogeneous overdensity. 
       The main result in our analysis is equations 
       (\ref{average-1}) and (\ref{average-2}).  
       The expressions indicate that the long-wavelength 
       inhomogeneity with the larger power might induce 
       the anisotropic flows and this could 
       dominate the expansion scalar and the shear tensor in the case 
       of the negative spectral index for the scale-free power spectrum. 
       Thus, the infalling velocity 
       around the averaged overdensity could be evaluated as the rather small  
       value, compared to the spherical infall model.  
       Solving the evolution equations
       (\ref{evolution1})-(\ref{evolution4}), 
       we have confirmed these things. 
       For $n<0$, the infalling velocity significantly 
       deviates from the prediction of the spherical infall model. 
       The same result is obtained in the case of the CDM spectrum 
       for which the dominant contribution of the background inhomogeneity 
       comes from the spectrum with the effective index $n_{\rm eff}<0$.  
       This indicates that we underestimate the 
      density parameter $\Omega$ when using the simple spherical infall
       model. It might lead to another suggestion that the 
       estimation of the cluster abundance using the Press-Schechter 
       formalism is changed when we take into account the effect 
       of the cosmological background density field. 
       On the other hand, for the spectral index $n>0$, 
       the effect of background inhomogeneity is negligible and 
       the simple spherical infall model could be a good 
       approximation for the infalling velocity. 

       The analysis of the infalling velocity with $n<0$ is 
       qualitatively consistent with the 
       numerical simulation of the clusters 
       of galaxies. Lee et al. (1986) analyzed the N-body model of a flat
       universe with an initial Poisson distribution and concluded
       that the prediction of the spherical infall model
       underestimates the density parameter in the mean though the
       scatter is large. Hoffman (1986) presented an analytical model
       of a spherical collapse with a global shear. He showed that
       when the ratio of the anisotropic flow to the isotropic infall 
       velocity exceeds about 50\%, infall velocities are higher than
        in the pure spherical infall model. In other 
       cases with the small anisotropy, corresponding to the situation 
       considered here, the simple dynamics (\ref{EOM-SCM}) 
       underpredicts the density parameter, which is consistent with our 
       result. Villumsen \& Davis (1986) studied the nature of the
       velocity field around large clusters in $\Omega =1$
       cosmological N-body simulations with more realistic initial
       conditions. They concluded that the substantial subclustering
       seen on small scales causes a systematic underestimate of the
       density parameter.   
       Lilje \& Lahav (1991) also considered the 
       averaged density and infall velocity profiles around clusters
       of galaxies in biased CDM models and showed that the prediction 
       of the spherical infall model overestimates the density 
       parameter in the case of the bias parameter 2.1. 
       As authors themselves stated, the several reasons could 
       cause the discrepancy: the selection and the definition 
       of the clusters, in connection with the normalization and 
       the bias parameter in the N-body simulation, and/or another 
       interpretation that the large-scale shear will make the infall 
       velocity larger. 
       It is therefore difficult to compare their result with ours. 
       Here we adopt the former reason which would be most 
       likely and conservative. Then, their result is reconciled with ours 
       provided that  the bias parameter is taken to be less than 1.8. 
       Thus, all of the other works could be in agreement with the present 
       analysis and provide us a simple physical picture: when the 
       spectral index $n<0$, the anisotropy of the random 
       velocities can act as an effective pressure and yield the 
       slowing down the infalling velocity.

       Although we could not find any simulations in the 
       cases with $n>0$, the similar behavior 
       can be found from the evolution of the power spectrum 
       (Makino, Sasaki \& Suto 1992, Jain \& Bertchinger 1994). 
       The weakly non-linear analysis using the perturbation theory shows   
       that the non-linear evolution of the power spectrum 
       has the different growth rate, depending 
       on the spectral index $n$. For the single power-law 
       spectrum $n<-1$, the non-linear growth of power 
       spectrum significantly enhances,   
       while the growth is suppressed if we have the index $n>-1$. 
       These behaviors remarkably coincide with our calculations of the 
       averaged quantities. Therefore, we might expect that our result 
       remains correct even in the $n>0$ case. 
       The validity of our prediction 
       will be confirmed by the numerical simulations elsewhere. 

       In this paper, we have found the significant contribution of the 
       background inhomogeneities to the infalling 
       flow around the overdensity. However, the influences of 
       cosmic background density field on 
       the local bound system deserve further study.  
       The background inhomogeneities also affects the internal 
       structure of the bound object. 
       The formation of substructure 
       near the central region significantly alters the dynamical  
       evolution. The non-radial motion induced by the tidal torques 
       might slow the dynamics of the collapse, which leads to the 
       previliarization process (Peebles 1990). 
       We think that the background density inhomogeneity could be 
       one of the most important sources of the previliarization 
       effect. To resolve this issue, the further analytical study 
       must be developed. 
%
%
%
%
\section*{Acknowledgments}
%
%
%
%
%
%
We are grateful to  M. Sakagami for valuable discussions and comments.  
A.T acknowledge the support of a JSPS Fellowship. 
J.S. is supported by the Grant-in-Aid for Scientific Research No. 10740118.
%
%
%
%
%
%
%
%
%

%
%
%
%
%
%
%
%
\appendix
%
%
%
%
%
%
\section[]{Calculation of the averaged quantities}
\label{sec: calculation}
%
%
%
%
In this appendix, using the definitions (\ref{average}) and 
(\ref{modified-P(k)}), 
       we calculate the averaged quantities 
       appearing in the right hand side of (\ref{R-evolution}) and 
       (\ref{rho-evolution}).

Using the new variables (\ref{separated-var}), 
       the averaged quantities can be written as 
\begin{equation}
{\bf C}_1 =~{\cal A},
\label{averagge-I}
\end{equation}
\begin{equation}
{\bf C}_2=-(4\pi)~{\cal A}-(4\pi)^2{\cal B}+{\cal C},
\label{averagge-II}
\end{equation}
where we used the relation (\ref{Poisson}). The quantities ${\cal A}, 
       {\cal B}$, and ${\cal C}$ are respectively given by 
\begin{eqnarray}
 {\cal A}= \langle\hspace*{-0.7mm}\langle
\partial_i\left[\hat{\delta}~\partial^i\hat{\Delta}\right]
\rangle\hspace*{-0.7mm}\rangle, 
~~~
 {\cal B}= \langle\hspace*{-0.7mm}\langle
\hat{\delta}^2
\rangle\hspace*{-0.7mm}\rangle, 
~~~
{\cal C}= \langle\hspace*{-0.7mm}\langle
(\partial_i\partial_j\hat{\Delta})(\partial^i\partial^j\hat{\Delta})
\rangle\hspace*{-0.7mm}\rangle,
\nonumber
\end{eqnarray}
where $\partial_i$ and $\partial^i$ are the derivative with respect to 
$x^i$ and the summation convention is used.  
%
In terms of the Fourier representation, the above equations become 
\begin{eqnarray}
 {\cal A}&=&-\frac{4\pi}{V_r}\int_{V_r} d^3x\int\frac{d^3k_1d^3k_2}{(2\pi)^6}
\left[1+
\frac{k_1^2+k_2^2}{2(k_1k_2)^2}(\k_1\cdot\k_2)
\right]
\nonumber \\
&&~~~~~~~~~~~~~~~~~~~~~\times
\langle\hat{\delta}(\k_1)\hat{\delta}(\k_2)\rangle~e^{-i(\k_1+\k_2)\x},
\label{A}
\end{eqnarray}
\begin{equation}
 {\cal B}=\frac{1}{V_r}\int_{V_r} d^3x\int\frac{d^3k_1d^3k_2}{(2\pi)^6}
~\langle\hat{\delta}(\k_1)\hat{\delta}(\k_2)\rangle~e^{-i(\k_1+\k_2)\x},
\label{B}
\end{equation}
\begin{eqnarray}
 {\cal C}&=&\frac{(4\pi)^2}{V_r}\int_{V_r} d^3x\int\frac{d^3k_1d^3k_2}{(2\pi)^6}
\left(\frac{\k_1\cdot\k_2}{k_1k_2}\right)^2
\nonumber\\
&&~~~~~~~~~~~~~~~~~\times~\langle\hat{\delta}(\k_1)\hat{\delta}(\k_2)\rangle
~e^{-i(\k_1+\k_2)\x}.
\label{C}
\end{eqnarray}
%
%
%
We first perform 
       the volume integral $V_r$ using the following formula : 
\begin{eqnarray}
 \int_{V_r}d^3x~e^{-i(\k_1+\k_2)\x}&=&(4\pi)^2~
\sum_{l=0}^{\infty}\sum_{m=-l}^{l}Y_{lm}^*(\Omega_1)Y_{lm}(\Omega_2)
\nonumber \\
&&~~~~\times ~\int_0^{r_0}dr~r^2j_l(k_1r)j_l(k_2r), 
\nonumber
\end{eqnarray}
where $Y_{lm}$ is the spherical harmonics and the argument $\Omega_i$ 
denotes the angular position for the wave vector $\k_i$. 

To proceed further, we rewrite the integral $\k_i$ with the 
spherical coordinates 
$(k_i, \theta_i, \phi_i)$. Then, we  
substitute the definition (\ref{modified-P(k)}) 
       into (\ref{A}), (\ref{B}) and (\ref{C}) and evaluate the 
       integral over $\theta_i$ and $\phi_i$. 
       For ${\cal A}$ and ${\cal C}$, 
       the calculation is tractable if we use the relations : 
\begin{eqnarray}
\left(\frac{\k_1\cdot\k_2}{k_1k_2}\right)
=\frac{4\pi}{3}\vec{Y}_1(\Omega_1)\cdot\vec{Y}_1(\Omega_2),
\nonumber
\end{eqnarray}
\begin{eqnarray}
\left(\frac{\k_1\cdot\k_2}{k_1k_2}\right)^2
 =\frac{4\pi}{3}\left[\vec{Y}_0(\Omega_1)\cdot\vec{Y}_0(\Omega_2)
+\frac{2}{5}\vec{Y}_2(\Omega_1)\cdot\vec{Y}_2(\Omega_2)\right],
\nonumber
\end{eqnarray}
where we define
\begin{eqnarray}
 \vec{Y}_l(\Omega_1)\cdot\vec{Y}_l(\Omega_2)\equiv\sum_{m=-l}^{l}
  Y_{lm}^*(\Omega_1)Y_{lm}(\Omega_2). 
\end{eqnarray}
With the careful manipulation, the following 
       results can be obtained:
\begin{eqnarray}
 {\cal A}&=&\frac{8}{V_r}\left\{\xi_0(r_0)-\xi_1(r_0)\right\}, 
\nonumber \\
 {\cal B}&=&\frac{2}{\pi~V_r} \sum_{l=1}^{\infty}(2l+1)\xi_l(r_0), 
\nonumber \\
 {\cal C}&=&\frac{32\pi}{V_r}
  \left[\sum_{l=0}^{\infty}(2l+1)\xi_l(r_0)-\frac{1}{3}\xi_0(r_0)-
\frac{2}{3}\xi_2(r_0)\right], 
\nonumber
\end{eqnarray}
where the quantity $\xi_l$ is defined by (\ref{xi}). Substituting 
the above expressions into (\ref{averagge-I}) and (\ref{averagge-II}), 
we finally get (\ref{average-1}) and (\ref{average-2}). 
%
%
%
%
%
\label{lastpage}
\end{document}